\documentstyle[epsfig]{aipproc}
\input epsf.tex
\def \gta {\mathrel{\vcenter
     {\hbox{$>$}\nointerlineskip\hbox{$\sim$}}}}
\def \m{\ifmmode M_\odot\else M$_\odot$\fi}
\def \r{\ifmmode R_\odot\else R$_\odot$\fi}
\def \l{\ifmmode L_\odot\else L$_\odot$\fi}

\def\gmcm3{gm~cm$^{-3}$}
\def\gm-s{gm~s$^{-1}$}
\def\cm3s{cm$^3$~s$^{-1}$}

\def\erg-s{erg~s$^{-1}$}

\def\beq{\begin{equation}}
\def\eeq{\end{equation}}
\def\ref{\reference}
\def\gr{$\gamma$-ray}
\def\grs{$\gamma$-rays}
\def\grb{$\gamma$-ray burst}
\def\grbs{$\gamma$-ray bursts}
 \def\co{$^{56}{\rm Co}\ $}      
\begin{document}

\title{Cosmic Explosions:\\Rapporteur Summary of the 10th
Maryland Astrophysics Conference}

\author{J. Craig Wheeler}

\address{Department of Astronomy\\The University of Texas\\
Austin, Texas 78751\\wheel@astro.as.utexas.edu}

\maketitle

\begin{abstract}

This meeting covered the range of cosmic explosions from solar
flares to \grbs.  A common theme is the role of
rotation and magnetic fields.  New information from the Sun
shows that a ``magnetic carpet" contains most of the surface
field that feeds the corona.  Disk instabilities in protostellar
disks may provide most of the growth of a protostar.  Type Ia
supernovae continue to give evidence for an accelerating
Universe, and a rigorous examination is underway to characterize
systematic effects that might alter the results.  The binary
evolution origin of Type Ia that explode as carbon/oxygen white
dwarfs at nearly the Chandrasekhar limit remains a thorny problem.
The discovery of the central point of X-ray emission in Cas A
by CXO should give new insight into the core collapse problem in
general and the nature of the still undetected compact remnant in
SN~1987A in particular.  Jets were described from protostars
to microquasars to blazars to \grbs.  Polarization studies of core-collapse
supernovae lead to the conclusion that core collapse is
not merely asymmetric, but strongly bi-polar.  To account for normal
core-collapse supernovae, the explosion must be jet-like in
routine circumstances, that is, in the formation of neutron stars,
not only for black holes. Given the observed asymmetries,
estimates of explosion energies based on spherically-symmetric
models must be regarded with caution.
The strong possibility that at least some \grbs\ arise
from massive stars means that it is no longer possible to decouple
models of the \grb\ and afterglow from considerations of the
``machine."  Although it began the
supernova/\grb\ connection, it is difficult to fit SN~1998bw/GRB~980425 into
any statistical picture that incorporates the high redshift events.
The implied correlation of
\grbs\ with star formation and massive stars does not distinguish a black hole
collapsar model from models based on the birth of a magnetar.
Current jet models do not discriminate the origin of the jet.
Calorimetry of at least one afterglow suggests that \grbs\ cannot
involve highly inefficient internal shock models.
Essentally all \grb\ models involve the ``Blandford Anxiety," the
origin of nearly equipartition magnetic fields in the associated
relativistic shocks.  It is important to discriminate the origin of
the \grb\ energy as thermal energy, kinetic energy or perhaps Poynting flux.

\end{abstract}

\section{Introduction}

The Tenth October Maryland conference celebrated the rapid
release of energy from objects ranging from protostars to
active galactic nuclei.  We were treated to an
inimitable history as only Virginia Trimble can do it from her
personal knowledge of the players and places and her voracious
reading of the literature.  Roger Blandford gave an overview
with a focus on unsolved problems.

In this summary, I will present a brief summary of the invited
review talks and selected poster presentations.
I regret, especially, that I cannot do
justice to all the latter.  To a certain extent, I will present
the great span of exploding objects we covered
as reflected in the mirror of supernovae.  There are three
reasons for this:  (i)~this is the subject I know best;
(ii)~various aspects of supernova research were presented at
length; and (iii)~supernovae, I believe, are related to
everything---and everything to supernovae.  The degrees of
separation are, in any case, rarely so great as six.
A more specific theme I will underline is that to an
ever-growing extent we are forced to consider rotation, magnetic fields,
and asymmetries.  In Section II, I will present some work
from my collaborators in Texas and elsewhere that relate to
this general theme.  Section III gives some perspective and
unresolved issues.

\section{Summary of Presentations}

\subsection{Stellar and Solar Flares}

Blandford emphasized that there has been great progress
recently in the study of the Sun that has led to a new and
different understanding of the magnetic field of the Sun and
its role in coronal heating.  In particular, solar flares come
from regions of relatively weak magnetic field where plasma can
intrude.  The magnetic field is higher in a ``magnetic
carpet" across the solar surfaces and it is from this
layer of stronger field that ``nanoflares" deliver energy to
power the solar corona.

Linsky reviewed our understanding of flares from a
variety of stellar sources.  These flares are of great
interest, but if other stars operate in
analogy to the Sun, then flares may be only the tip of
otherwise unseen magnetic activity in other stars.  Ramaty
emphasized the manner in which flares can be used to study
the acceleration of particles.  Kenyon outlined the great
progress that has been made in terms of understanding the FU
Ori phenomenon as a disk instability in a protostellar disk,
pointing out that all the growth of a protostar may arise
during the phases of high luminosity and high mass accretion
rate.  Clearly, study of these phenomena, especially in
the detail afforded by the Sun, can give us valuable insight into the
processes of reconnection and coronal formation that are, in
turn, crucial in order to understand accretion disks, winds,
jets, and particle acceleration.

\subsection{Type Ia Supernovae}

The use of Type Ia supernovae as ``calibrated candles'' by
means of empirical brightness-decline relations has been
startlingly successful.  The tentative conclusion that there
is a low matter density and a finite, positive cosmological
constant has sent reverberations throughout astronomy and
physics (Riess et al. 1998; Perlmutter, et al. 1999).
Type Ia are more complex than can
be described by the first versions of one-parameter
brightness decline relations: $\Delta$M(15) (Hamuy, et al. 1996), Multicolor
Light Curve Shape (Riess, Press, \& Krishner 1996), or stretch
(Perlmutter et al. 1999).  Theorists can, and have,
invented many reasons why Type Ia might vary with look back
times.  On the other hand, the first order corrections have
been remarkably successful in reducing the scatter in the
data and providing constraints on cosmological parameters.
In this conference, Kirshner and Aldering presented data
showing that no significant evolutionary effects have yet
appeared in the data.  In particular, Aldering argued that
refined analysis of error bars yields no statistically
significant evidence that the early rise times are different
in nearby events and those at redshift z$\sim$0.5
(e.g. Reiss et al. 2000).

Nomoto outlined models for Type Ia evolution and possible
systematic effects.  Of particular interest is the recent
calculation of H\"oflich et al. (1999) showing that, at lower
metallicity, a star of a given mass will produce a smaller
C/O ratio.  All else being the same (ignition density,
density of transition to detonation), a lower C/O ratio
will lead to a somewhat brighter event with a faster
decline.  The one parameter brightness-decline relations
would then interpret such an event as somewhat dim, just the
effect interpreted as evidence for a cosmological constant.
More modelling of this sort and comparison with data is
necessary to elucidate these sorts of physical systematic
effects.

There were also discussions of possible sources of observational
systematic bias.  Suntzeff gave an excellent review of the observations
and a cautionary note about comparing photometric results from
different observatories that use slightly different filters sets
and thus get slightly different results for standard stars in
standard filters.  There is a tendency for the deep searches
to not follow up events that are at larger distance from potential
host galaxies because of the ambiguity of association of the
supernovae candidate with the host.  Filippenko pointed out that
there is some complementary bias against candidates that are close
to the centers of host galaxies because those are sometimes passed
over when classification spectra are obtained due to concerns about
galactic contamination.  These selection biases in terms of
the radial position on the galaxy may in turn be important at
some level because the distribution of intrinsic luminosity and
decline rate is known to vary with galactocentric radius
(Wang, H\"oflich \& Wheeler 1997; Riess et al. 1999; Howell, Wang
\& Wheeler 1999).  A bias against events at low galactocentric
radius might give a bias toward events that are more homogeneous
in their properties, but less able to discriminate subtle
differences in progenitor dependence.  The success of the light curve
brightness/decline relations to remove potential evolutionary effects
by including sample events that span the full range of progenitor ages
and metallicity in nearby events in spirals and ellipticals might
thus be subtly affected.

My personal answer to the question of whether the Universe is
acclerating is ``probably yes."  My answer to the query of
do we know for sure is ``not yet."

The physics of Type Ia
supernovae was discussed, without which there will remain some doubt concerning
the purely empirical treatment of Type Ia as cosmological probes.
Pinto gave a summary of the basic understanding of why
there is a brightness/decline relation.  He noted that the
observed relations can be produced in white dwarfs of constant
mass, near the Chandrasekhar mass in the most successful models.
In this context, if the nickel mass is increased the peak
luminosity naturally goes up.  Less obvious is that the temperature
goes up resulting in a increase in the line opacity. This traps
the energy from \gr\ decay, resulting in a slower decline, the
process amply illustrated in previously published models
(H\"oflich Khokhlov \& Wheeler 1995; H\"oflich \& Khokhlov 1996;
H\"oflich et al. 1996).

The physical question that emerges from such an analysis
is why the nickel mass should vary when the white dwarf progenitor
mass is essentially fixed.  This is presumably
a product of initial conditions and the thermonuclear
combustion of the white dwarf.  The physics of the combustion was
described by Hillebrandt and by Niemeyer.  There has been
great progress in recent years in applying concepts from
terrestrial combustion physics to
the Type Ia problem.  Observations and theory suggest that
the combustion begins with a relatively slow, subsonic,
turbulent deflagration.  To account for the distribution of
intermediate mass elements in velocity in the outer layers of
Type Ia explosions, there must be a transition to a much faster burning.
The issue of whether there is a natural transition from subsonic
deflagration to supersonic detonation has been discussed
(Khokhlov, Oran \& Wheeler 1997a,b,c; Niemeyer \& Woosely 1997;
Khokhlov, Oran, Wheeler \& Chtchelkanova 1999; Montgomery, Khokhlov,
\& Oran 1998).  Hillebrandt and Niemeyer raised the question
of the difficulty of making a direct transition from deflagration
to detonation and discussed the possibility that a speed up
to a very rapid, but still subsonic deflagration was possible
and adequate to account for the observations.  It is not clear
that a very rapid deflagration would not be unstable to evolution
to a detonation.  One thing that is clear is that the turbulent
deflagration must be studied in three dimensions to get the
sign of the turbulent cascade (from large scales to small
scales) correct and to understand that process in the context
of spherical dilution and related effects (Khokhlov 1995).

Another major issue that arises in the context of the standard,
Chandrasekhar mass model for Type Ia supernovae is the
question of their prior evolution.  How do white dwarfs
grow to the Chandrasekhar mass sufficiently often to account
for the observed rates of Type Ia?  Nomoto described the models
that currently seem to come closest to solving this long-standing
issue.  These models invoke a wind from the white dwarf so that
a relatively rapid mass transfer rate from the companion does not
glut the white dwarf with mass to yield a surrounding hydrogen-rich
envelope, in violation of the observations.  Rather, the excess
mass can be blown off in the wind.  Models show that the net
accretion onto the white dwarf can be rapid enough to avoid
degenerate hydrogen or helium ignition.  Such ignition is inimical
to the Type Ia process since a nova explosion will reduce the
mass of the white dwarf, as discussed by Hernanz.  Degenerate
helium ignition produces a supernova explosion of the wrong properties,
as outlined by Nomoto.  In the wind models, the amount of hydrogen
on the surface of the white dwarf when it explodes is sufficiently
small to escape detection.  For a given wind model, there are
solutions that will give the desired properties of Type Ia
progenitors with unstable mass loss from main sequence companions
and stable mass loss on the thermal time scale from red giant
companions with masses in the range 1 - 3 \m.
At lower metallicities, the wind could be less
efficient and it may not be possible to produce Type Ia.  This
would give an epoch of turn-on of Type Ia with look back time.

One of the issues raised by these binary wind models is the
reservoir problem.  The amount of mass that is lost from the
secondary is related to the mass that accretes  onto the white
dwarf, promoting its growth. This can be expressed by:
\[
\Delta M_2 = \Delta M_{wd}\left(1 + \frac{\dot M_{wind}}{\dot M_{wd}}\right).
\]
If the secondary only has a mass of 1 - 3 \m, the wind model may
work if the rate of loss to the wind is comparable to the rate of
accretion onto the white dwarf.  The companion loses only two
grams for every gram that lands on the white dwarf.  On the other
hand, wind mass loss rates are not known very well in most
circumstances, certainly not in this rather exotic one, and factors
of a few may be important.  If, for instance, the rate of loss to
the wind is several times the growth rate of the white dwarf, then
if the white dwarf must accrete several tenths of a solar mass to
reach the Chandrasekhar limit, the small mass companion might
not be able to provide enough.  This, of course, depends on the
initial mass of the white dwarf.  If the supernova progenitor
must grow from the mass of a field white dwarf, 0.6 \m, the
task is nearly insurmountable.  If the initial mass of the white dwarf
is substantially higher, the task is easier.  An important issue then
becomes the initial mass distribution of the white dwarfs, a
function that is not well known at higher white dwarf masses and
which is undoubtedly affected by the very condition of being in
a binary system.  Hernanz and Starrfield pointed out that there
are white dwarfs in binary systems with rather large masses,
for instance that in the recurrent nova system U Sco at
about 1.3 \m, so nature can do this, at least occasionally.

It is still a high priority to obtain any information that will
give us hints of the nature of the binary system underlying
Type Ia supernovae.  If they are hydrogen accretors, as the
binary wind models suppose, the explosion should take place
next to a hydrogen-rich companion and within the wind.  The
search for the stripped companion is important, but handicapped
because the hydrogen tends to lag the ejecta and is expected to
show up, if at all, in the nebular phase when the spectrum
is complex and weak H$\alpha$ might be difficult to detect.
Suntzeff pointed out that the use of sensitive eschelle detectors
on the new generation of large aperature telecopes might
give a new way to search for the hydrogen swept up in the
wind.

\subsection{Collapse}

The process of core collapse is of great current interest both
for its intrinsic importance as a supernova triggering mechanism
and for its potential connection to \grbs.  Fryer gave an
update on work to understand collapse, especially
the difficult problem of neutrino transport and the critical
issue of fall back which may determine whether a neutron star
or black hole is left behind in a successful supernova explosion.
Fallback considerations suggest that black hole formation could be
common in stars with mass as low as 20 \m, making SN 1987A right
on the ragged edge of going either way, neutron star or black hole.
Fryer predicted that after a decade of concentration on
multidimensional hydrodynamics and specifically protoneutron
star convection, the next decade would be one devoted to
the effects of rotation and neutrino transport.  I think the
former is unambiguously true and would add magnetic fields
to the mix.  As for the latter, it is clear the neutrinos will
continue to play a large role since they must carry off the
bulk of the binding energy of the neutron star.  It is not
so clear that they will emerge as the final arbitrating physics
of the success or failure of the explosion itself as the
role of jets becomes more clear, as discussed in \S D.

A great new step toward understanding the outcome of core collapse
came with the launch of the new Chandra Observatory.  The first
obtained and released image of Cas A represented an incredible
debut.  Although there were hints of a central object in ROSAT
data, the Chandra image showed with unambiguous clarity the
dim point of X-ray emission in the center of the remnant.  The issue
of whether this is a cooling neutron star or an accreting
neutron star or black hole is now under debate.  One thing is
clear.  This image gives us a new slant on similar issues in
SN 1987A where the expected neutron star has still not revealed itself.
If the object left behind in SN 1987A is similar to that in Cas A,
then it is no wonder we have not seen it.  The object in Cas A is
very faint, less than a few \l, perhaps as little as
$10^{32}$ \erg-s (Tanenbaum, et al. 1999).
One possibility to detect the central object in SN 1987A is to
register the bolometric emission from the absorption and re-emission
of any source within the ejecta.  Heroic efforts to measure the
bolometric luminosity still place it at more than $10^3$ \l, more
than 1000 times brighter than the object in Cas A.  It may be that
by searching diligently in the continuum between emission lines
tighter limits can be obtained in SN 1987A, but detecting an object
as faint as that in Cas A will be a challenge.  
Figuring out the nature of the object in Cas A will immediately
give us new perspectives on SN 1987A.  One of the issues will
be to understand why this object is $10^4$ to $10^5$ times dimmer
than the 1000 year old pulsar in the Crab nebula.  If neutron
stars with rapid rotation and strong magnetic fields are necessary to
make jets in supernovae, Cas A and SN 1987A do not seem to qualify.
There is, however,  obvious evidence for some jet-like activity
in Cas A and SN 1987A has its rings and asymmetric ejecta, so this
story has a long way to run.

To take a step back, Davidson regaled us with his recriminations that we
have worked so little on, and understood so little of, $\eta$ Carinae.
He is exactly right.  We have made a lot of progress modeling
massive stars as spherically symmetric, but somewhere we may have
taken a drastically wrong turn since we are very far from predicting
the properties of $\eta$ Carinae from first principles.  The lack
of progess in understanding $\eta$ Carinae is not due to lack of
interst or curiousity, but simply a lack of knowledge of where
to start in this complex beast.  Surely we must try.  The mass of
the star is likely to exceed 100 \m, and so it must surely collapse
or explode.  It might be a \grb\ waiting to happen.  The lobes and
skirt are compelling in their beauty and simplicity and strongly
reminiscent of the rings of SN 1987A.  The details of ropes, strings,
and jets are bewildering, as is the great outburst of the last
century and the recent brightening.  Once again the Chandra image
of Cas A brings a new vision of this key object.

There is also much to entertain after the explosion.  McCray
outlined the three-ring circus that is beginning to ensue with
the first lighting of the first blob in the inner ring
of SN~1987A as the fastest
ejecta collide with the most prominent protrusions.  McCray
revealed that the ``knot" that has lit up may not have the
enhanced N abundance reflecting CNO processing that is ascribed to
the rings, but rather may be more representative of the ISM
of the LMC.  This would be an intriguing result.  In any case,
the next decade will be a three-ring circus replete with
fireworks as the ejecta continue to interact with the ring.
There is much to be learned about interstellar medium shock
physics as well as the nature of the progenitor star and
the ejecta.

\subsection{Jets}

The topic of jets has long been of central interest to astrophysics.
There is a new concentration on relativistic jets because of their
possible role in \grbs.

Livio gave an overview, pointing out that jets are ubiquitious,
from protostars to \grbs.  He suggested that they
have a common mechanism powered by accretion, with the ejection velocity
being comparable to the escape velocity from the surface of
the central star or the inner edge of a surrounding accretion disk
and the collimation by a surrounding magnetic field.  This may
even apply to jets from the nuclei of planetary nebulae, although
it is not clear what the source of accreted matter is in that
context.  Possible jets from pulsars may represent an exception to
this general mechanism.  Again the images of the Crab pulsar from
CXO show a jet-like protrusion that may help our understanding
of these issues.

Another lesson is that the jets, especially those associated with
black holes, are frequently relativistic.  This was emphasized in
the reviews of jets from blazars by Urry and of those from the
binary black holes in ``microquasars" by Greiner.  Studies of these
relativistic jets on both galactic and stellar scales with
time-dependent, multiwavelength campaigns has great potential to
teach us about how the jets form, are collimated, and propagate.
The stellar cases are especially important because the activity
plays out over a shorter timescale and is especially amenable to
practical, detailed study.

Some of the greatest interest in jets is the possibility that
they play a role in the origin of \grbs.  Jets are
discussed as a way of moderating the great energy requirements of
the brightest bursts and there is some circumstantial evidence for
collimation in the change of slope of some afterglow light curves
(Rhoads, 1999; Sari, Piran, \& Halpern 1999; Stanek, et al. 1999;
Harrison et al. 1999 ).
On the theoretical side, the models of Woosely, MacFadyen, and their
collaborators have drawn great interest.
These models are computed
in the context of a ``collapsar" model where a black hole forms
by gravitational collapse in the center of a massive star.  Rotation
could lead to the formation of a disk surrounding the black hole.
The disk, in turn, could generate strong neutrino fluxes that
might provide an energy source by neutrino/antineutrino annihilation,
or it could be the source of an MHD flow.  Yet another alternative is
that the spin of the black hole threaded by magnetic fields could
generate energy by the Blandford-Znajek effect (Blandford \& Znajek 1977).
In the models
discussed by MacFadyen \& Woosley (1999), energy from neutrino annihilation
is presumed to be deposited as thermal energy in a small region
along the rotation axis.  In their two-dimensional simulation, the
expansion from this point is channeled by the density gradient
of the surrounding Keplarian disk and is forced to proceed up the
rotation axis.  As the density at the jet base declines, thermal
energy input at a given rate tends to provide an ever larger
specific energy to the matter.  The result is to promote the
formation of a relativistic jet that is confined and collimated
by the structure of the star.  As emphasized by Woosley, the
jet will trigger lateral shocks that can cause the outer mantle
to explode.  Woosley differentiated possible differences
between situations involving production of the black hole by
prompt collapse or by fallback.

The production of
something like a supernova attendant to the propagation of the
jet through the stellar core seems unavoidable.
The jets made in this way can, in principle, be relativistic,
and can, again in principle, yield \grbs.  The open issues
are whether collapse leads to jets, the nature of those
jets, and the question of whether the jets will lead to
\grbs\ of observable properties.

\subsection{Gamma-Ray Bursts}

Fishman summarized the history of the \grb\ game and especially
the invaluable role of BATSE on CGRO.  BATSE had discovered
2612 bursts at the time of the meeting and added one more
the evening the meeting ended.   The smallest time resolved
in a \grb\ is 200 microseconds.  If interpreted in terms of
an intrinsic light crossing time, that corresponds to a distance
of less than 60 km.  Fishman argued that despite some reports
to the contrary, the short bursts are not homogeneous.  They
display, for instance, $V/V_{max} = 0.39$, significantly less
than the homogeneous value of 0.5.  There are also claims for
anisotropy, but Fishman did not think those were well substantiated
by the data.

Kulkarni summarized the recent history in the BeppoSAX era.
Of special interest to \grb\ research and to this conference in
particular are the recent reports by Bloom et al. (1999),
Reichert (1999) and Galama et al. (1999) for supernova-like modulation
of \grb\ afterglow light curves about three weeks after the \grb\
for two cases of classic \grbs.  In the thinking of many people,
this increases the already high probability
that GRB~980425 was associated with SN~1998bw (Galama, et al. 1998).  
GRB~980425 does not, however, fit in the context of various statistical
studies of, e.g. Schmidt (1999) on luminosity functions,
of Ruiz \& Fenimore (1999) on correlations of variability with
luminosity, and Norris (1999) on energy-dependent phase lags.
If SN~1998bw and GRB~980425 were the same event, the \gr\
emission process was very different than the more distant events.

Kulkarni reviewed the valuable radio data that has been obtained
on afterglows.  He pointed out that the failure to detect radio
afterglows in some cases may simply be because the existing
equipment, e.g. the VLA, is not sufficiently sensitive.  He
also reported on the calorimetry of one event, GRB~970508, for
which late time radio observations could be modeled to obtain
an estimate of the total energy radiated in the afterglow 
(Frail, Waxman \& Kulkarni 1999).
The result was $E_{afterglow}/E_{\gamma} << 1.$  This is significant
because the most popular models of internal shocks imply low
efficiency (Kumar 1999) and hence a great deal more total initial kinetic
energy in relativistic baryons (in synchrotron models) than
in emitted $\gamma$-rays.  The
kinetic energy left from the initial \grb\ must all be dissipated
in the subsequent interaction with the ISM and hence radiated in
the afterglow.  Taken at face value, this result does not
support the inefficient internal shock model.  The calorimetry
depends on assuming synchrotron emission and hence nearly
equipartition magnetic fields.  The issue of how, and hence
whether, equipartion fields arise in relativistic blast waves
is very unclear, as emphasized at this conference by Blandford.
More calorimetric information of this kind is clearly needed.
Another object that gives some information of this sort is
GRB~990123, the famous bright prompt optical burst (Akerloff et al. 1999;
Kehoe et al. 1999; Kulkarni et al. 1999).
The total fluence in the optical of that burst was substantially
less than the isotropic equivalent fluence in \grs.  On the
other hand, a backward extrapolation of the optical afterglow intersects
at a point above the prompt flash, so the
relative balance of \grb\ to afterglow energy is uncertain.
Kulkarni also raised the issue of whether the afterglow must
be non-adiabatic in contrast with popular models.  This
could complicate the analysis and alter the energetics and
hence the basic constraints on the processes of the \grbs\
and afterglows.  There are models that avoid the problem
of inefficient internal shocks by invoking a collision of the
leading shock with the external medium
(Fenimire \& Ruiz 1999) or by ``spotty" internal shocks
(Kumar \& Piran 1999).

Kulkarni gave a brief summary of what he termed as the
indirect indications for a correlation of \grbs\ with
massive stars, including the location in host galaxies
and apparent correlation with star forming regions.
He concluded that the data, indirect though it is, points
to a collapsar model.  This conclusion may be a bit too
specific.  The evidence points to a correlation with massive,
short lived stars, but it contains no direct information
on the specifics of the mass of the stars and certainly not
whether the event involves the formation of a black hole.
To be specific, the data are equally consistent with a massive
star that makes a magnetar - a rapidly rotating, highly magnetized
neutron star.  Theory suggests that magnetars must be born
rapidly rotating and must dump a great deal of rotational
energy to slow to the long periods observed 10,000 years later
in the soft \gr\ repeaters.

Gehrels outlined the exciting future for observations of \grbs.
HETE II is scheduled for launch January 23, 2000
from Kwajalein Island.  HETE II should bring the era of afterglow
studies from short as well as long bursts and should produce a
much higher rate of well-localized \grbs\ than BeppoSAX.  This
will make life for the observers doing ground-based follow-up even
more hectic.  SWIFT was selected as a MIDEX instrument just after
the meeting.  It is scheduled for launch in perhaps 2003.  A great
deal of work and discovery on \grbs\ will occur between now and
then, but there should be much left for SWIFT to do.  In particular,
the prospect of extending the detection to events at redshift of
10 to 20 is extremely exciting for cosmology as well as \grb\ research.

M\'esz\'aros summarized the theory of \grbs\ and their afterglows,
so much of which he pioneered before the discovery of afterglows.
He remarked that this progress was possible in part because of
the fortunate circumstance that the physics of the internal shock
region and that of the external shock/afterglow region can
be decoupled from the physics of the ``machine," the process/object
that actually produces the energy that is transformed into
relativistic shocks and \grs.  The field has developed to the point
where, increasingly, this may no longer be true.  For instance,
depending on the choice of Lorentz factor and energy of the
burst, the radius of the photosphere of the fireball could be
less than the radius of the bare helium core that is invoked in
jet models (Khokhlov et al. 1999; MacFadyen \& Woosely 1999).
In addition, there are issues, again arising in the context of
massive star models, of winds.  These high density winds will
change the length scale of interactions to produce prompt
optical output in reverse shocks and subsequent afterglows by
external shocks.  The era is upon us when we must begin to
consider the machine self-consistently with the \grb\ and afterglow.

\subsection{Type I X-ray Bursts}

Bildsten and Swank summarized the progress made on understanding
the Type I X-ray bursts, especially with the invaluable
contribution of RXTE which has allowed new insight based both
on new data and new understanding of old data.   The Type I
X-ray bursts are complementary to many of the other objects
discussed here in other contexts.  Unlike the magnetars
in the soft \gr\ repeaters, for instance, which are highly
magnetized and rather slowly rotating, the neutron stars
associated with Type I X-ray bursts apparently have fast
rotation and low magnetic fields.  RXTE data of pulses from
one source is interpreted as evidence for rotation at 300 Hz.
Bildsten described the systematics of nuclear burning on
the surfaces of neutron stars and interpreted the data as
evidence for a hot spot ignited by a localized thermonuclear flash
that is whipped around by the rotation.  The lesson that fits in
with the general theme of this summary is that the Type I
X-ray bursts require asymmetry, rotation, and magnetic
fields. Bildsten also outlined the possibility that nuclear
burning in this ambiance could break out of the hot CNO and
helium burning and run all the way up to heavy elements,
including, for instance, species like krypton.\footnote{This
caused me to pose the following question:  if Type I bursts
make krypton(ite), did Jor-el and his son come from a
planet orbiting a neutron star?}

\section{Polarization and Jets in Normal Core-Collapse Supernovae}

To complement the theme of asymmetry, rotation and magnetic fields,
I would like to summarize some work that we have done at Texas
over the last five years.
Like many people in the supernova community, we at the University
of Texas got actively involved in the supernova/soft-gamma-ray
repeater/magnetar/\grb\ topic with the advent of SN~1998bw and
its possible connection to GRB~980425.  We brought a different
perspective to this issue because of work we have done 
on supernova spectropolarimetry.

We have been making spectropolarimetric observations of all
accessible supernovae at McDonald Observatory (Wang et al. 1996;
Wang, Wheeler \& H\"oflich 1997; Wheeler, H\"oflich \& Wang 1999;
Wang et al. 1999). A summary of observations is given in Table 1.
The result has been that most Type Ia have low polarization and
hence are substantially spherically symmetric.  Many have only
upper limits of order 0.1 - 0.2\%.  A few have detected, but
low polarization, of order 0.2\%.
We have obtained the first polarization of a
``subluminous" Type Ia, SN~1999by which appears to show
polarization at the 0.2\% level.
The polarization observed
is consistent with theoretical models of delayed detonation
models (Wang, Wheeler \& H\"oflich 1997) and may be a useful
probe of the combustion physics.  We have detected one exception,
SN~1997bp, which was observed a week before maximum light to
have a polarization of about 1\%.  The polarization was low in
post-maximum spectra, but this event remains a challenge to
understand.
It is important to establish whether such events are
common, the physical reason for the large polarization, and
whether or not there could be an asymmetric luminosity distribution
that could affect estimates of cosmological parameters.

More importantly in the current context are our observations
of presumed core-collapse events, Type II and Type Ib/c.  We have
found that all such events are polarized at about the 1\% level
and some much more so.
So far there have been no exceptions in about a dozen events
(a recent Type II, SN~199em, showed no detectable polarization in 
very early observations (Leonard, Filippenko \& Chornock 1999), 
but further observations are planned that will
peer deeper into the ejecta).
There could be a myriad reasons for polarization, but our data
suggest a very important trend: the smaller the hydrogen envelope,
the larger the observed polarization.  As examples of this trend, SN~1987A
with a 10\m\ envelope had a polarization of about 0.5\% ( M\'endez et al.
1988); SN~1993J with a small hydrogen envelope, $\sim0.1$\m, was polarized
at the 1-2\% level (Trammell, Hines \& Wheeler 1993; Tran et al. 1997);
a very similar object, SN~1996cb, may show polarization as high as 4\%;
Type Ic SN~1997X which showed no substantial hydrogen nor helium was
polarized at perhaps greater than 3\% (Wheeler, H\"oflich \& Wang 1999);
SN 1998S which shows characteristics of a Wolf-Rayet star (Leonard et al.
1999; Gerardy et al. 1999) showed polarization of about 3\% before
maximum (Leonard et al. 1999) and perhaps as much as 4\% after maximum
(Wang, et al. 1999).

\begin{center}
Table 1. Supernovae with polarimetric measurements\\
\begin{tabular}{llll|llll}\\
\hline
\hline
\label{wheeler-tab1}

SN       &  Type &  $P$(\%) & Intrinsic &     SN  &  Type & $P$(\%) &
Intrinsic\\
\hline
SN1968L$^1$ &  II   &  0.2      & No        & SN1970G$^2$ &
II   &  0.5    &    Yes\\
SN1972E$^3$ &  Ia   &
0.35    & No        & SN1975N$^4$ &  Ia   &  1.5    &
No\\
SN1981B$^5$ &  Ia   &   0.41    & No        & SN1983G$^{6,7}$ &  Ia   &
2.0    &    No\\
SN1983N$^{6,7}$ &  Ib   &           & Yes?     & SN1987A$^{8,9}$ &  II   &
0.5    &
Yes\\  SN1992A$^{10}$ &  Ia   &
0.3      & No        &SN1993J$^{11,12}$ &  IIb  &  1.5    &
Yes\\
SN1994D$^{12}$ &  Ia   &  0.3      & No
& SN1994Y$^{12}$ &  II   & 1.5     &    Yes\\
SN1994ae$^{12}$&  Ia   &  0.3      & No        & SN1995D$^{12}$ &
Ia   & 0.2     &    No\\
SN1995H$^{12}$ &  II   &
1.0      & Yes       & SN1995V$^{12}$ &  II   & 1.5     &
Yes\\
SN1996W$^{12}$ &  II   &  0.7      & Yes       & SN1996X$^{12}$ &  Ia   &
0.2     &    Yes?\\
SN1996cb$^{12}$&  IIb  &  3.0      & Yes       & SN1997X$^{12}$ &  Ic   &
2-7?
&    Yes\\
SN1997Y$^{12}$ &  Ia   &  $<$0.3   & No
&SN1997bp$^{12}$ &  Ia   & 1.0     &    Yes\\
SN1997bq$^{12}$&  Ia   &  $<$0.2   & No        & SN1997br$^{12}$&
Ia   & $<$0.2  &    No\\
SN1997ef$^{12}$&  Ic? &  $<$0.3
& No        &SN1997ei$^{12}$ &  Ic   & 2.5     &    Yes\\
SN1998S$^{12,15}$ &  IIn  &  3.0      & Yes &SN1998bw$^{13,14}$ &  Ic? &
0.4?  & Yes?\\
SN1999by$^{12}$ &  Ia
&  0.2    &    Yes\\
\hline
\hline
\end{tabular}
\end{center}
1 Wood \& Andrews (1974), 2 Shakhovskoi \& Efimov (1973),
3 Wolstencroft \& Kemp (1972), 4 Shakhovskoi (1976),
5 Shapiro \& Sutherland (1982), 6 McCall et al. (1984), 7 McCall (1985),
8 Cropper et al. (1988), 9 M\'endez et al. (1988),
10 Spyromilio and Bailey (1993), 11 Trammell, Hines \& Wheeler (1993),
12 This program, 13 Key et al. (1998), 14 Patat et al. (1998),
15 Leonard et al. (1999)\\[2mm]

These are difficult observations requiring
special care in the reduction to remove the effects of the ISM
(the latter greatly aided by wavelength and temporal coverage).
Following Suntzeff's cautionary notes on the difficulty of doing
accurate photometry, Leonard referred to spectropolarimetry as
``photometry from hell."  In addition,
there is a pressing need to expand the statistical sample,
especially with time-sampled data.
Nevertheless, this trend suggests that the core-collapse process
itself is strongly asymmetric and that evidence for that asymmetry
is damped by the addition of outer envelope material.

The level of polarization we have observed for core collapse events,
$\sim1$\%, requires a substantial asymmetry with axis ratios
of order 2 to 1 (H\"oflich 1995).
Asymmetric explosions tend to turn spherical as
they expand, so to leave a significant imprint
in the homologously expanding matter requires a substantially larger
asymmetric input of energy or momentum in the explosion process
itself (H\"oflich, Wheeler \& Wang 1999).
In other words, the asymmetries we are observing require
the underlying explosion to be driven by a jet.  This conclusion
is completely independent of any connection to \grbs, but, of
course, the potential for this connection is clear
(Wang \& Wheeler 1998; Wheeler 1999).
These factors led us to the hypothesis that the core collapse
process is intrinsically strongly asymmetric, much more so than
current collapse calculations involving convectively unstable
neutron stars.  It was in this context that we greeted the
news of SN~1998bw and have continued to work on polarization,
jet models of collapse, and their possible relation to other
astrophysical phenomena.

\subsection{SN~1998S}

SN~1998S was discovered on March 2, 1998.  It showed strong
narrow emission lines and is thus characterized as a Type IIn.
Some of the narrow emission lines were of high excitation,
reminiscent of Wolf-Rayet stars and SN~1983K (Niemela et al. 1985)
and subsequently
showed emission of carbon monoxide and possible evidence for
dust formation that are consistent with an origin in the core of a
massive star that is not decelerated by a substantial hydrogen
envelope (Gerardy et al 1999).  Leonard et al. (1999) were very
fortunate to be at Keck II with a spectropolarimeter and got an
excellent set of data on March 7, still about 0.5 magnitude
and two weeks before maximum.  They have interpreted their
data in terms of an interaction with a disk of circumstellar
hydrogen which also shows up as a double (in fact triple) peaked
H$\alpha$ profile at late times.  The polarization might
have been as high as 3\%.

The light curve declined
rather rapidly after maximum unlike some Type IIn, so it
has been called Type IIn(pec).  About 60 days after the explosion
(20 days after maximum) it went into a steeper decline and then
leveled off to a slower decline about 80 days after maximum.
This slower decline is, in V, still a little steeper than expected for
\co\ decay.\footnote{http://oir.www.harvard.edu/cfa/oir/Research/supernova/spectra}
We obtained data on SN~1998S at McDonald Observatory
on March 31, about 10 days after maximum and again on May 1,
60 days after the explosion, 40 days after maximum and just
before the light curve started to decline rapidly.  The McDonald
data is consistent with that of Leonard et al. following  a locus
nearly parallel (but perhaps somewhat shifted) in the Q,U plane.
The polarization on May 1 could be as high as 4\%.

Polarization of this level forces us to abandon more timid phrases
like ``asymmetric supernovae."  Recall that the maximum polarization
from an infinitely thin, internally illuminated, electron scattering
disk is about 12\% (Chandrasekhar 1950).  On this scale a polarization
of 4\% is very large.  For this event, and perhaps for others with
this level of polarization, it is appropriate to talk about
``bi-polar supernovae," not merely asymmetric supernovae with
the implication, perhaps, of irregularities rather than large,
well-ordered imprints of basically asymmetric geometry.  In principle
an infinitely thin, internally illuminated rod could be 100\% polarized,
but we do not think this geometry corresponds to the observations
and the dynamics of the events.

\subsection{Jets and Magnetars}

To explain the polarization data of routine core collapse
supernovae, we need an explosion mechanism with a stong,
indeed, bi-polar asymmetry that can survive the dynamics
of expansion and remain substantial in the homologous
phase.  We need a jet.  To account for normal
supernovae we must have jets in routine circumstances, that
is, the formation of a neutron star and not restricted to
the more rare circumstances of the possible formation of
a black hole.  This statement is independent of
the liklihood that in rare cases or different circumstances
such a jet might yield a \grb.

The obvious place to look for jets in frequent core
collapse events is in the rotating,
magnetic collapse of a neutron star with the equivalent dipole
magnetic field ranging from ``typical" values like the Crab
pulsar to the extreme values associated with magnetars
and soft \gr\ repeaters (Kouveliotou et al. 1998).
This environment gives a framework in which to
quantitatively address questions of physics that are
germane to the nature of the core collapse process in general and to
potential \gr\ production.
The physics that could be at play
in such a collapse has recently been considered by Wheeler
et al. (1999).

Rotation and magnetic fields have a strong potential to create
axial matter-dominated jets that will drive strongly asymmetric
explosions for which there is already ample observational
evidence in Type II and Type Ib/c supernovae, their remants,
and in the pulsar velocity distribution.
The potential to also create strong flows of Poynting flux
and large amplitude electromagnetic waves (LAEW)
serves to reinforce the possibility to generate bi-polar explosions.
These bi-polar explosions will, in turn, affect nucleosynthesis and issues such
as fall-back that determine the final outcome to leave behind
neutron stars or black holes. In addition, the presence of
matter-dominated and radiation-dominated jets might lead
to bursts of \grs\ of various strengths.  The issue of the
nature of the birth of a ``magnetar" in a supernova explosion
is of great interest independent of any connection to \grbs.
Highly magnetized neutron stars might represent one out of
ten pulsar births.  Production of a strong \grb\ is probably 
even more rare.

Wheeler et al. (1999) show that the contraction phase of a proto-neutron star
could result in a substantial change in the physical properties
of the environment.
When the rotating, magnetized neutron star first forms there is
likely to be linear amplification of the magnetic field and
the creation of a matter-dominated jet, perhaps catalyzed by MHD effects,
up the rotation axis.
The rotational energy of the proto-neutron star is typically
about $10^{51}$ ergs.  The energy
of the proto-neutron star is sufficient to power a significant
matter jet, but unlikely to generate a strong \grb.  The matter jet could
generate a smaller \grb\ as seems to be associated with
SN~1998bw and GRB~980425 by the Colgate (1974) shock acceleration mechanism as
it emerges and drives a shock down the stellar density gradient
in the absence of a hydrogen envelope, e.g., in a Type Ib/c supernova.

As the neutron star cools, contracts,
and speeds up, two significant things happen.  One is that
the rotational energy increases.  The energy becomes significantly
larger than required to produce a supernova and sufficient,
in principle, to drive a cosmic \grb\ if the collimation
is tight enough and losses are small enough.
For a neutron star with a period near 1 millisec the rotation
energy can be substantially in excess of $10^{52}$ ergs.
The rotational energy of the contracted neutron star
is radiated away in the form of a Poynting flux or LAEW at the
frequency $\Omega_{NS}$.  If efficiently utilized and
collimated, this energy reservoir could make a substantial
\grb.   The luminosity is estimated to be
\[
L_{EM}\simeq4\pi R_{LC}^2\times\frac{c}{4\pi}|\vec E\times\vec B|
\simeq \frac{\mu_{NS}^2}{R_{LC}^4}c
\simeq\frac{R_{NS}^6B_{NS}^2 \Omega_{NS}^4}{c^3},
\]
assuming the LAEW to be generated at $R_{LC}$ and the magnetic moment
of the neutron star to be $\mu_{NS}=B_{NS}R_{NS}^3$. For the conditions
of the contracted neutron star which has initiated an $\alpha-\Omega$
dynamo, we expect
\[
L_{EM}\simeq4\times10^{52}~{\rm erg~s^{-1}}
\left(\frac{R_{NS}}{10~{\rm km}}\right)^6
\left(\frac{B_{NS}}{10^{16}~{\rm G}}
\right)^2\left(\frac{\Omega_{NS}}{10^4~{\rm s^{-1}}}\right)^4,
\]
which will last for a duration of several seconds.

The second important factor the accompanies the contraction
and spin-up of the cooling neutron star is that the light cylinder contracts
significantly, so that a stationary dipole field cannot form
and the emission of strong LAEW occurs.  Tight collimation
of the original matter jet and of the subsequent flow of LAEW
in a radiation-dominated jet is expected.

The LAEW will propagate as intense low frequency, long
wavelength radiation.  
The LAEW ``bubble" could be strongly Rayleigh-Tayor unstable,
but still may propagate selectively with small opening angle
up the rotation axis as an LAEW jet.  Alternatively, the
impulsive production of LAEW could render the stellar matter
nearly irrelevant as a confining medium.
If a LAEW jet forms, it can drive shocks which may selectively propagate
down the axis of the initial matter jet or around the perimeter of
the matter jet.  The shocks associated with the LAEW jet could generate
\grs\ by the Colgate mechanism as they propagate down
the density gradient at the tip of the jet or there
could be bulk acceleration of protons to above
the pion production threshold.  The protons could produce
copious pions upon collision with the surrounding wind,
thus triggering a cascade of high energy \grs, pairs,
and lower-energy \grs\ in an observable \grb.
Yet another alternative is that the LAEW
could eventually propagate into such a low density
environment that they directly induce pair cascade.

The matter-dominated jet
requires 5 to 10 seconds to reach the surface of the neutron
star, just about the time for the neutron star to cool, spin
up and launch the second, faster, more energetic jet.
The second, LAEW-driven jet propagating out at nearly the speed of
light could thus arrive at the surface of a bare helium core at
just about the time of the earlier MHD jet launched when the
protostar forms, but which propagates more slowly.
The natural time scale for
any \grb\ is about 5 to 10 sec, the cooling, spin down time
for the neutron star, but shorter times scales could be
associated with the shock breakout, and instabilites in
the LAEW production process or in the flow.
The question of what fraction of the pulsar energy
goes to drive quasi-spherical expansion and what
fraction propagates as co-linear LAEW clearly
requires greater study.

Issues of uncertain
physics aside, it is clear that this mechanism might
not be robust in the production of \grbs, but might
produce \grbs\ of varying strength depending on
natural variation in the circumstances of a given
collapse event.
Any \grs\ emitted by any of these processs
are likely to be strongly collimated.
The luminosity of the emitted radiation will depend on the
geometry of that emission.  The energy
produced by the spin-down of the pulsar could emerge from
the stellar surface along the axis of a low-density matter jet, or in
an annulus surrounding a high density jet.  Either of
these cases will give a Lorentz factor that depends strongly
on the aspect angle of the observer.  Computation of the
resulting luminosity is thus distinctly non-trivial.

\subsection{Jets and Bi-Polar Supernovae}

It remains to be proven that newly formed neutron stars can produce
jets.  In the meantime, one can
study the dynamics of jets and their impact on the stars in
which they are generated.   A preliminary study in which
conditions were selected to represent the sort of MHD jet
found by LeBlanc \& Wilson (1970; see also M\"uller \& Hillebrandt 1979;
Symbalisty 1984) has been presented by Khokhlov et al (1999).
This study has been extended to explore a range of jet energies
and stellar configurations, both bare helium cores and
red supergiants.

The code developed by Khokhlov (1998) is an Eulerian
adaptive mesh code based on the Piecewise Parabolic Method.  The
calculations are fully three dimensional.  The adaptive mesh gives
excellent resolution.  The finest scale corresponds to a uniform grid
of some $10^{10}$ cells.  The adaptive mesh also allows great
dynamic range. For the jet models this ranges from $2^{12} \sim 10^4$ to
$2^{19} \sim 10^6$.  The imposed jets are cylindrically symmetric
and the initial stellar model is spherical.  The resulting jets are
thus highly cylindrically symmetric, but this is not imposed in
the dynamics, only the initial conditions.  The jet dynamics are
sufficiently rapid for the models computed that Kelvin/Helmholz
instabilities have little time to form.

Figure 1 shows the distribution of the jet matter (unspecified in the
computation, but presumably rich in iron-peak elements), and of
the oxygen layers of the star.  The former reflects the bi-polar nature
of the jet flow. The latter shows the effects of the lateral shocks that
compress the oxygen into an equatorial shell.  This will, in turn,
affect the line profiles of the oxygen observed in the
nebular phase.  These profiles are presented after 4.84 seconds when
the jet breaks through the surface of the helium core.  They must
be followed into homologous expansion before any direct connections
to observations can be made.

\begin{figure}[t]
\begin{minipage}[t]{2.5truein}
\mbox{}\\
\epsfig{file=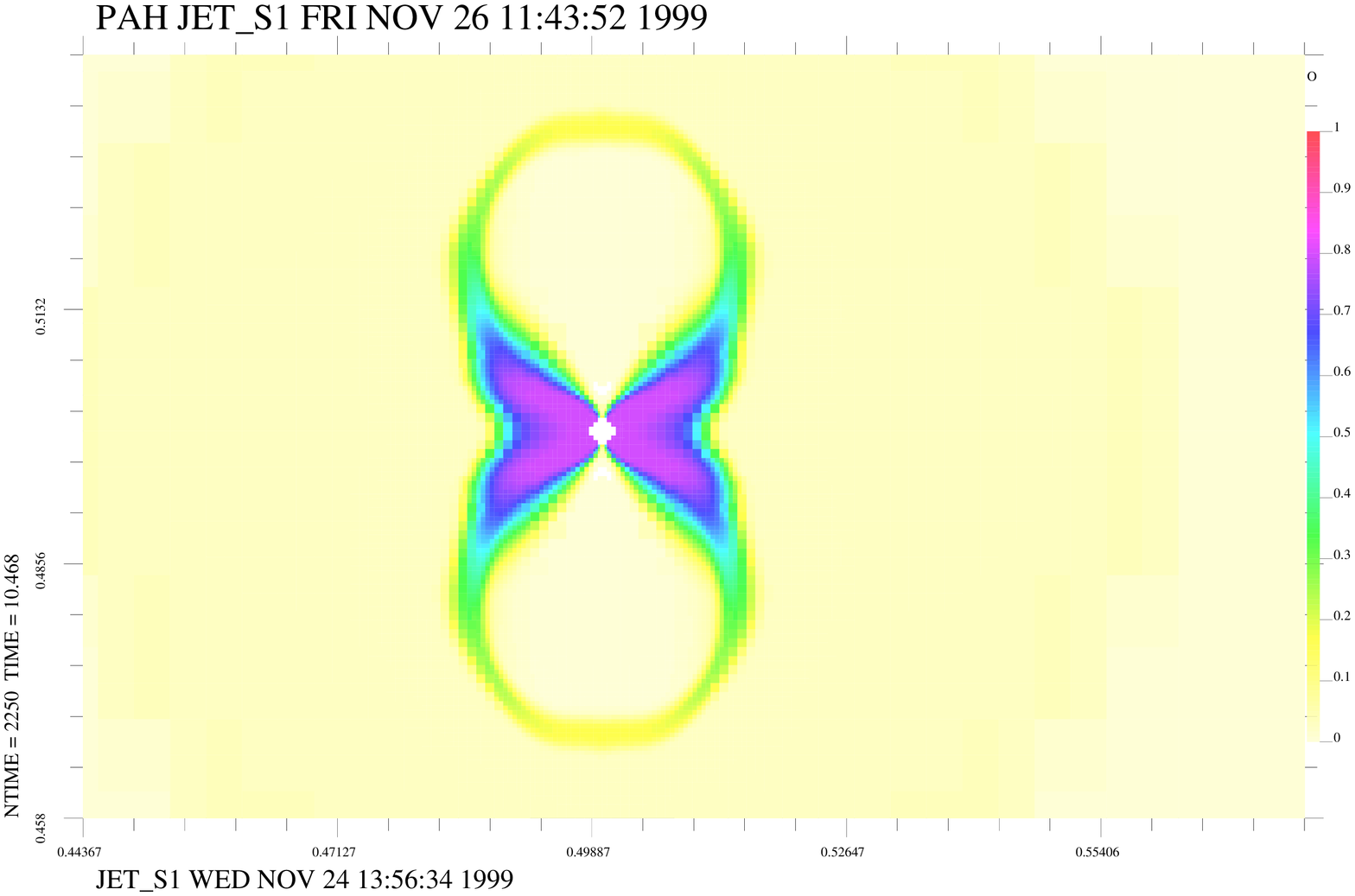,height=2.5truein,width=2.5truein,
clip=,angle=270}
\end{minipage}
\hfill
\begin{minipage}[t]{2.5truein}
\mbox{}\\
\caption{
Composition structure of a jet-driven supernova.  The axial jet
(light lobes) contains jet material, presumably rich in iron-peak
material.  The equatorial
shell (darker region) shows the distribution of the oxygen layer from
the initially spherical progenitor model (from H\"oflich et al. 2000).
}
\end{minipage}
\label{figure 1}
\end{figure}                               

We have also studied models with red giant envelopes.  The code
allows us to follow the jet in a single calculation from the
center of the star out through the extended envelope.  We find
that energetic jets can penetrate the hydrogen envelope, but
that more modest jets cannot.  The latter can still induce
an asymmetric, bi-polar explosion.

\section{Issues and Perspectives}

I will dwell in this final section mostly on issues of the
supernova/\grb\ connection since that is a topic of such
excitement and the one highest on my personal interest scale.

The drive to understand \grbs\ must address issues of inhomogeneity.
Is the mechananism related to massive stars or not?  Is the machine
related to neutron stars or black holes or both?  When are winds
important, when only the ISM as a working surface and catalyst for
external shocks?  The Lorentz factor $\Gamma(t, \theta)$ is almost
surely a function of time and angle.  
Another important issue is the degree of collimation.
A break in the afterglow light curve can signify that a relativistic
jet has slowed to subrelativistic speeds and is beginning to expand
laterally.  Alternatively, the interaction of a blast wave with
a dense wind can produce qualitatively the same result.  
This spirit of inhomogeneity
has been nicely captured by Chevalier \& Li (1999) who have analyzed which
events are most likely to have occurred in massive star winds and
which have not (see also Frail et al. 1999).  Table 2
gives a compilation of the results of Chevalier \& Li.

\begin{center}
Table 2. Gamma-Ray Bursts and Afterglows\\
\begin{tabular}{lllll}\\
\hline
\hline
\label{wheeler-tab2}
Burst & Redshift & Afterglow Type & Supernova & Jet\\\hline

970228 & 0.695 & wind & yes & \\

970508 & 0.835 & wind & no  & no \\

980326 &  & wind & yes & \\

980425 & 0.0085 & wind & SN~1998bw & no \\

980519 & & wind & ?  & \\

990123 & 1.60  & ISM & & yes \\

990510 & 1.619 & ISM & no  & yes \\\hline\hline

\end{tabular}

\end{center}

One of the most interesting issues is whether \grbs\ arise in
neutron stars or black holes, or both.  
Black holes almost certainly exist.  They are
observed in galactic cores and in binary X-ray sources.
In addition, black holes make relativistic jets.  This is
again seen in both active galactic nuclei and in the binary
X-ray sources, especially the microquasars.  
We also know neutron stars exist, and the soft-gamma ray repeaters
have provided evidence that magnetars 
exist.  The magnetars may have dipole fields of $10^{14}$ Gauss or more,
substantially above the limit where the magnetic field
affects quantum electrodynamics.  The question of the nature of
the birth event of a magnetar is clearly an important one,
independent of issues of connections to \grbs.  Another
important fact is that all core-collapse supernovae are
polarized and that there is growing evidence that the explosion
must be, not just irregular, but bi-polar.  By demographics, this
must apply to events that form neutron stars, not just the more rare
events associated with black hole formation.

One issue is then what this circumstantial evidence is telling
us about the nature of neutron stars and black holes and their
possible relation to \grbs.  Table 3 gives some features of
the astrophysical events we are attempting to relate, as
discussed by various speakers at this meeting.
Urry noted that blazars never drop below $1\%$ of the peak flux during
fluctuations, whereas some \grbs\ have gaps with no detectable flux.
Livio characterized jets as arising from conditions with
$v_{jet}/v_{esc}\sim 1$, a condition clearly related to the value
of $\Gamma$ in the jet.

\begin{center}
Table 3. Properties of Black Hole Systems and Gamma-Ray Bursts\\
\begin{tabular}{lllll}\\
\hline
\hline
\label{wheeler-tab3}
Object & $\Gamma$ & L/L$_{Edd}$ & L$_{interpulse}$ & $v_{jet}/v_{esc}$ \\\hline

blazar & $\sim$ 10 - 20 & $\sim 1$ & $\gta 1 \%$ & $\sim 1$ \\

microquasar & $\sim 10$ & $\sim 1$ & ?? & $\sim 1$ \\

\grb\ & $>$ 100 & $\gta$ $10^{12}$ &  sometimes $\sim 0$ & c - $\epsilon$
\\\hline\hline

\end{tabular}

\end{center}

One of the suggestions from Table 3 is that, left to their own
devices, black holes produce relativistic jets, but they do not,
in the context of AGNs and microquasars, produce the highly
relativistic flows thought to occur in \grbs.  This may mean
that black holes  cannot produce \grbs\, or it may mean that
the circumstance of a black hole in a \grb\ system must be
substantially different.  One possibility for the
latter is that the black hole can not be in a relatively
isolated environment, but must, for instance, be surrounded by
baryons, by a star, to help focus and amplify the flow to
very high Lorentz factors.  Another point is that there is some
tendency to think that the canonical \gr\ energy of a \grb\ is about
$10^{52}$ ergs with higher apparent energies being due to
collimation.  If this is so, then the possibility of a neutron
star generator is still alive.  Possibilities are the collapse
of an iron core and the birth of a magnetar, accretion induced
collapse, or the merger of two white dwarfs.  On the other hand,
if the production of \grs\ is inefficient and substantially
more than $10^{52}$ ergs of total energy is required, then
the possibility of a neutron star progenitor will die.

SN~1998bw continues to play a large role in the on-going debate
concerning the supernova/\grb\ connection.  A comparison of
the properties of ``normal" hydrogen and helium deficient Type Ic
supernovae and the peculiar SN~1998bw is instructive. Type Ic are polarized.
SN~1998bw was polarized.  Type Ic probably require a jet-like flow
of energy and matter to produce a bi-polar explosion.  So does
SN~1998bw.  Routine Type Ic presumably leave behind neutrons
stars.  The speculation is that SN~1998bw left a neutron star or
a black hole.  The current evidence is mute on which.  Routine
Type Ic require about $10^{51}$ ergs of kinetic energy.  SN~1998bw
requires $\gta 10^{52}$ ergs if the explosion was spherical.  We
know the explosion was not spherical, so this energy estimate is
somewhere on the continuum from uncertain to misleading to wrong.
If the explosion produced a photosphere with a 2 to 1 axis ratio,
then, with proper aspect of angle of the observer, it might
require only $\gta 10^{51}$ ergs (H\"oflich, Wheeler \& Wang 1999).  
One can also argue that this value is on the same continuum from
uncertain to misleading to wrong until quantitative non-spherically
symmetric radiative transfer is done.  Even if energies in
excess of $10^{52}$ ergs are required for SN~1998bw and other events,
this does not necessarily mean they made black holes.  This
energy is certainly in the range that could come from tapping
the rotational energy of a new neutron star.

Another important issue is to begin to consider the \grb\ and
afterglow mechanisms self-consistently in the context of
the ``central machine" and its environment.  Specifically,
there are issues that must be faced in contemplating an
origin of \grbs\ in massive stars.  The star is there, and so,
presumably, is the dense wind such stars are expected to shed.
Consideration of the star and wind will affect the development
of the \grb\ in a relativistic ``impulse" or ``wind" and
the density of the environment must affect the length and
time scales over which the afterglows are produced.

The production of shocks and radiation surely depend on
the manner in which the energy is delivered.  In the
currently most popular model for \grbs\ and afterglows,
a large kinetic energy is produced in a relativistic wind
that either has irregularities imposed on it from the
``machine" or develops irregularities by instability
in the flow.  These irregularities produce ``internal shocks"
and the \grs.  The residual kinetic energy
propagates into the external medium to
produce an ``external" shock and the afterglow.  A possible
alternative is that energy is delivered, for instance at
the outer boundary of a helium core rather than from
near the last stable circular orbit of an isolated black hole,
by a strong Poynting flux.  This Poynting flux could lead,
via pair formation, directly to \grs.  The residual energy of
the pairs would then be needed to power the afterglow.  The
difference in these two processes might be substantial.
For instance, Usov (1999) has pointed out that in a strong
Poynting flux, internal shocks are impossible because there
can be no relative motion of the advected particles.  Another
issue as he expressed it in this meeting is the ``Blandford
Anxiety," the origin of the nearly equipartition magnetic
fields that are invoked and/or derived in these models.
An intense Poynting flux could deliver the magnetic field
directly to the environment where a  \grb\ is triggered,
if not in the larger region of the afterglow.

The bottom line is that the near future of the study of
cosmic explosions, like the recent past, is likely to
continue to make our heads spin.  The objects we study not
only spin, they are magnetic and asymmetric.  Coping with
that should lead to great insight and progress.

\begin{acknowledgments}

I am grateful to the organizers for the invitation to the
meeting and an excellent scientific program
and to all the speakers for holding my attention
through the whole conference.  Special thanks
go to the staff for ensuring that the meeting arrangements went
smoothly and to the students who toted the microphones
around the hall so we could all be heard.  I am
especially grateful to my colleagues Peter H\"oflich,
Lifan Wang, Alexei Khokhlov, and Elaine Oran who
have taught me so much about explosions on and off the Earth.
This research was supported in part by NSF Grant
9818960 and a grant from the Texas Advanced Research
Program.

\end{acknowledgments}

\end{document}